\documentstyle[prl,aps,twocolumn,epsf]{revtex}
\def\break#1{\pagebreak \vspace*{#1}}
\begin{document}
\draft
\title{Eulerian Walkers as a model of Self-Organised Criticality}
\author{V. B. Priezzhev$^{1}$, Deepak Dhar$^{2}$, Abhishek Dhar$^{2}$ and Supriya Krishnamurthy$^{2}$}
\address{$^1$ Laboratory of Theoretical Physics, Joint Institute for Nuclear
Research, Dubna, Moscow region, 141980 Russia \\
$^2$Theoretical Physics Group, Tata Institute of Fundamental Research,
Homi Bhabha Road, Bombay 400005, India}

\date{\today}

\maketitle
\widetext
\begin{abstract}
We propose a new model of self-organized criticality.
A particle is dropped at random on a lattice and moves along
directions specified by arrows at each site. As it moves, it changes
the direction of the arrows according to fixed rules. On closed graphs
these walks generate Euler circuits. On open graphs, the particle
eventually leaves the system, and a new particle is then added.
The operators corresponding to particle addition
generate an abelian group, same as the group for
the Abelian Sandpile model on the graph. We determine the
critical steady state and some critical exponents exactly, using this
equivalence.
\end{abstract}


\narrowtext

In recent years, there has been much interest in the study
of systems showing self-organized criticality (SOC) \cite{bak} and
different models have been proposed for many systems such as
sandpiles \cite{bak},
earthquakes \cite{chen}, forest-fires \cite{dros} and
biological evolution \cite{baksn}. All these models
involve a slowly driven system, in which the externally-introduced
disturbance propagates in a random medium using deterministic or
stochastic rules. In the process it modifies the medium so that after
many such disturbances, the medium develops long-range spatial correlations
\cite{fly}.

The most analytically tractable of all these models has
been the Abelian sandpile model (ASM) \cite{deepak,prez}.
In this Letter we introduce a new model of SOC called the Eulerian
walkers model (EWM). This model is quite different from the ASM in
some ways, but shares with it the abelian group property. This allows
a determination of the steady state and exact calculation of some
critical exponents. In fact we define a general abelian model of which
both the EWM and ASM are special cases.

 In the EWM self-organization
occurs due to activity of a walker which moves deterministically in a
medium while also modifying it. We show that on  closed
graphs, the walker finally settles into a limit cycle which is an Euler
circuit visiting each directed bond exactly once in one cycle. On open
graphs, the particle eventually leaves the system. We then add at a
randomly chosen site another
particle, which then moves, and so on.  We
define particle addition operators which act
on the set of recurrent configurations of the system. These operators
generate an abelian group, and satisfy same closure relations between
themselves as in the ASM on the same
graph. We show that recurrent configurations of the system are in one
to one correspondence with spanning trees on the lattice.

The model is defined for a general graph as follows:
consider a connected oriented graph $G$ consisting of $N$
points $i=1,2...N$. A point $j$ has $\tau_j$ outgoing bonds, and an equal
number of incoming bonds, connecting it to other points.
The outgoing bonds at $j$ are
\break{1.3in}
labelled by integers from $1$ to
$\tau_j$. We associate with each point, an arrow which can point along
one of the outgoing bonds (Fig. 1). Let $n_j$ $(1 \leq n_j \leq
\tau_j)$  denote
the current direction of the arrow that is the label of the bond along
which the arrow points. The set $\{ n_j \}$ specifies the arrow
directions at all points and provides a complete description of the
arrow configuration of the medium .

We now put a walker at some point on the graph. At each time step: \\
$(i)$ the walker after arriving at a site $j$ changes the arrow direction
from  $n_j$ to $n_j+1$(mod$\tau_j$),  \\
$(ii)$ the walker moves one step from j along the new arrow
direction at j.

Thus the motion of the walker is deterministic, is affected by the medium
and in turn affects the medium .
We can interpret the rules (i),(ii) as an intention of the walker to
maximize intervals between successive visits of the same bond each time
the walker leaves a given site.
The current position of the walker along with
the value of the variable $n_j$ at every site $j$ specifies completely
the state of the system \cite{lang}.

In the absence of sinks the walker continues
to walk forever. Since the system (walker+medium) has a finite number
of possible states, it eventually settles into a limit cycle. In
general, one would expect the size of these
cycles to be of the order of Poincare recurrence
times for the system, and grow exponentially
with N. The surprising fact is that all the cycles are very short
and, in fact, are of the same length $\sum_{j=1}^N {\tau}_j$. In each
such cycle, all 
bonds are visited exactly once (Fig. 1). Such walks are
known as Euler circuits \cite{har} and their study
 has been an important problem in
lattice statistics (If the circuit visits all {\it{sites}} exactly once, it
is called a {\it{Hamilton}} circuit).
There is a one to
one correspondence between Euler circuits and spanning trees on the
same graph. Clearly for any Euler walk ending at a site $j$, last exit
bonds from all
sites other than $j$ form a spanning tree rooted at $j$. Kasteleyn
also showed that each rooted spanning tree corresponds to a unique
Euler circuit \cite{kas}. The number of
all possible trees is known to be given in terms of the
determinant of the adjacency matrix by the well known matrix tree
theorem \cite{kas}.

We now show that every limit cycle is an Euler circuit.
We start from some arbitrary initial state of the medium with the
walker at some point $i$. The walker leaves the point $i$ along some
bond $b_1$. We evolve it till after time $T$ it returns to $b_1$ for
the first time. Let the bond traversed at the $j$th step of path be
$b_j$, so that the path is $b_1b_2...b_T$ with $ b_{T+1}=b_1$. We
can show that no other bond in this path is visited twice.
Proof: Assume the contrary and suppose that during the $T$ steps the bond $c$,
originating from the point $j$, is the first bond that is visited
twice. Each successive exit from $j$ is along a different direction so
there will be $\tau_j+1$ exits. But the number of visits to $j$ equals
number of exits. Hence there must exist some bond going into $j$
which is also passed more than once. This contradicts the fact that
$c$ was taken to be the first bond to be passed twice. Thus all $b_i$,
$i=1$ to $T$ are distinct and hence $T \leq \sum_{i=1}^N {\tau_i}$. 
If every bond in $G$ is visited we have an Euler circuit with
$T=\sum_{i=1}^N {\tau}_i$. If not,
consider the path $b_2b_3...b_{T+1}b_{T+2}$. If $b_{T+2}=b_2$ we
have another circuit of length $T$. We keep shifting the path thus 
till we reach a $t$ such that $b_{T+t} \neq b_t$. Such a $t<T$
exists so long as there are points $j$ on the path
which have not been visited $\tau_j$ number of times. Let $T'$ be the
first time when $b_{T'+t}=b_t$. Clearly $T<T'$. Now define the new
circuit formed by the $T'$ steps starting with the $t$th step. 
Iterating this we get circuits of increasing
lengths $T<T'<T''...$ where each is $\leq \sum_{i=1}^N {\tau}_i$ and so
finally we will get an Euler circuit when $T=\sum_{i=1}^N {\tau}_i$.
All the configurations which the system 
goes through before it enters the cycle are transients.

To illustrate the process of self-organization,
consider the motion of a walker on an infinite line starting
with a random initial configuration of the  medium. This walk has a
simple structure (Fig. 2). 
The walker turns the arrow at the origin and starts the motion
along the new direction of this arrow  reversing the arrows at all
sites it passes through. It moves on till it encounters a site
with an arrow pointing in the direction of motion. The walker now
reverses its direction and retraces its path entirely, passing over all
the sites traversed since the last reversal of its direction. Then it
continues to move ahead till it again encounters an arrow 
pointing in the direction of motion and so on.
Thus the arrows in the region already visited get organized
into an almost Eulerian circuit so that, if at time $t$ the number of
sites visited is $S(t)$, then in the previous $2S(t)$ time steps,
most of these sites have been visited exactly $2$ times. In
addition, the boundary of the cluster advances by a finite amount
$\Delta$, as some new sites are visited. 
For compact clusters 
$ S(t) \sim R(t)$, the average distance of the of the walker from the
origin, at time $t$. Thus we get
\begin{equation}
 \frac{dR(t)}{dt} \sim \frac{\Delta}{R} ~.
\end{equation}
which implies that $R \sim t^{\frac{1}{2}}$ for large $t$. The
average number of sites visited till time $t$, $S(t)$, goes as
$t^{\frac{1}{2}}$.

In higher dimensions, the motion of the EW is not so simple.In Fig. 3
we show the results of a simulation of the model on a square
lattice with random initial configuration of arrows. Before each step
the arrow is turned clockwise by $90^{o}$. The sites visited at least
once by the walker form a cluster with few holes, whose radius $R(t)$
increases with time $t$. In the region visited by the walker, all
arrows are not aligned parallel, but are organized into an almost
Euler circuit so that in the time between $T$ and $T-4S(T)$, only a
very small fraction of sites is ${\it{not}}$ visited ${\it{exactly}}$
$4$ times (here $\tau_j=4$ for all sites $j$).
Arguing as in the one dimensional case we get $R \sim t^{\frac{1}{3}}$
for large $t$. We have carried out Monte-carlo simulations and
verified this to very good accuracy.

However, for $d>2$, a random walker does not return to previously visited sites
often enough, and we expect the motion of a walker in an initially
random medium to be diffusive ($R^2 \sim t$). Our numerical
simulations show that this is indeed the case for $d=3$.

Now consider an open graph for which all the external perimeter sites
are identified with a single sink site, $i_0$, at which the
walker gets absorbed. We place a walker at some point $i$, with
probability $p_i$ ($\sum{p_i}=1$),
and let it evolve according to the rules specified before, until it
leaves the system [The walker will not get into a cycle as every cycle
would contain all points of the graph including $i_0$].
Now the system is specified only by the values of
{$n_i$}, $i=1,N$. We define operators $a_i$
acting on the space of recurrent configurations of the EWM as follows:
for any recurrent configuration $C$, $a_iC=C'$, where $C'$ is the
resulting configuration of the medium obtained by adding a particle at
site $i$ on the configuration $C$, and evolving it until it leaves the system.

It is easy to see that the operators at different sites commute.
Treat the motion of each walker as a sequence of elementary steps.
Then if two particles (walkers) are added to the lattice at sites $j$
and $j^{'}$, the
elementary moves on two sites  $j \not = j^{'}$ commute. If $j
= j^{'}$, they also commute due to identity of particles.
Therefore
\begin{equation}
[a_i,a_j] =0 ~.
\end{equation}
Within the space of recurrent configurations the operators $a_i$ will
have unique inverses.
If we define the $N \times N$ matrix, $\Delta$, such that $\Delta_{ii}$
gives the number of outgoing bonds from $i$ and $- \Delta_{ij}$ gives the
number of bonds from $i$ to $j$ then
\begin{equation}
\prod_j a_i^{\Delta_{ij}}=I~~\rm{,~for~all~}i
\end{equation}
which simply reflects the fact that $\tau_i$ particles added at i produce
the same effect as 1 particle added at each nearest neighbor
of $i$.
Thus, operators $a_i$ satisfy the
same algebra as the particle addition operators in the ASM.
In fact one can define
commuting operators $a_{i}(\epsilon)$, which have a phase factor
proportional to the number of steps taken by the walker in going from
$C$ to $C^{'}$ as in the case of the ASM \cite{dhar}.

So, as in the ASM, the number of recurrent
configurations $R=Det(\Delta)$ and they occur with equal probability.
The first result also follows from the one to one correspondence between
steady state configurations of the EWM and spanning trees which is
obtained by drawing the last exit directions at all points.
Since the $a_i$'s commute, they can be diagonalized simultaneously.
Then as for the ASM Eq.(3) determines all the eigenvalues of $a$'s.
Thus one can diagonalize the evolution operator $W=\sum p_i a_i$,
where $p_i$ is the probability that a new particle is added at site
$i$. For a lattice of size $L$ on a d-dimensional lattice, this
implies that the largest relaxation time of the system varies as
$L^d$.

Let $G_{ij}$ be the expected number of full rotations of the arrow at
the site $j$ due to addition of a walker at the site $i$.
During the walk, the expected number of
steps leaving $j$ is $G_{ij} \Delta_{jj}$ whereas $-\sum_{k
\not=j}G_{ik}\Delta_{kj}$ is the average flux into $j$. Equating
both fluxes one gets
$$\sum\limits_{k}G_{ik} \Delta_{kj} = \delta_{ij}~~~~~~\rm{or} $$
\begin{equation}
G_{ij} = [ \Delta^{-1} ]_{ij}.
\end{equation}
Average number of steps $n$ taken by the walker till
it leaves the system is given by $<n>=z<s>_{ASM}$, where $s$ is the
number of topplings in ASM avalanches,  for regular graphs
with coordination number $z$. Hence $<n> \sim L^2$, where $L$ is the
length of the system.

It is quite straightforward to calculate the arrow-arrow correlation
function in the steady state using the equivalence of the problem to
spanning trees. For two given sites $\vec{R_1}$ and $\vec{R_2}$ the
probabilities that arrows at the sites are in the directions
$\vec{e_1}$ and $\vec{e_2}$ respectively is the ratio of spanning
trees with these bonds occupied to the number of all spanning trees.
This is easily calculated. For large $R_{12}$ the leading term
in the connected part of this probability is given by
\begin{equation}
C(\vec{R_{12}};\vec{e_1}\rm{,}\vec{e_2}) \sim
(\vec{e_1}.\vec{\nabla}{\phi(\vec{R_{12}}}))
(\vec{e_2}.\vec{\nabla}{\phi(\vec{R_{12}}}))
\end{equation}
where
$\phi(\vec{R_{12}})=G_{\vec{R_1}\vec{R_2}}-G_{\vec{R_1}\vec{R_1}}
$. In $d$ dimensions $\phi(R)$  varies as $R^{2-d}$, hence the
correlation function $C(R)$ varies as $R^{2-2d}$ for large separations
$R$ \cite{note}. Thus the steady state of the model has
long range correlations and hence exhibits self-organized
criticality.

     As pointed out earlier, when the walker has left the system, the   
medium is in a recurrent state, and the arrows form a spanning tree.
This is not true for intermediate times where the motion of the EW may 
lead to a cyclic configuration of arrows. Thus a typical evolution of 
medium has periods of cyclicity interspersed between `normal' acyclic
states. In the EWM, the durations of these intervals of cyclicity
have a power-law distribution. In two dimensions, numerical simulations
\cite{shch} show that the probability of intervals of cyclicity of
duration $\tau$ varies approximately as $1/\tau^{1.75}$.

Though we can establish a one to one correspondence between the recurrent
configurations of the EWM and ASM, the relaxation process in the
two models is quite different. In the latter (in more than one
dimensions) in almost all cases
particle addition leads to
a stable configuration after a finite number of topplings, and the fraction
of avalanches which reach the boundary is very small.
In contrast, in the present model, each walker must travel to the
boundary before it
leaves the system, and thus the fraction of events involving  a finite
number of steps of the walker is zero in the limit of large system
sizes.  This leads to the interesting conclusion that the statistics
of avalanches is not completely determined by the operator algebra of ASM.

However, in one dimension, the probability
distribution of number of steps can be computed exactly, and we find
that apart from trivial numerical factors, it is exactly the same form
as the limiting distribution found by Ruelle and Sen \cite{ruse} for
avalanches in the $1$d ASM.

To bring out the relationship of the present model to the ASM more
clearly, we observe that due to the abelian nature of the evolution
rules, we can add and evolve two or more walkers in the system in
arbitrary order without effecting the final state. Let us choose the
following rules: each walker arriving at a site waits there until the
number of particles waiting at that site is $\ge$ r. Then these r particles take 1
step each in the directions $n_j+1$,$n_j+2$...$n_j+r$, and the arrow
is reset to $n_j+r$(mod$\tau_j)$. Clearly $r=1$ corresponds to the
EWM, and $r=\tau_j$ corresponds to the ASM. In the latter case the arrow
configuration does not evolve at all, and may be omitted from
discussion.In Fig. 4, we have
shown the results of Monte Carlo simulation of this general model on
a square lattice of size $200 \times 200$ for $r=1$ to $4$.
We see that we get the same general behaviour of distribution of
avalanches for all $r > 1$, but the case $r=1$ is special. It belongs to
a different universality class. For small $s$, the distribution $P(s)$ is
dominated by boundary avalanches, therefore, it does not have a simple
thermodynamic limit. However, the model has long range correlations,
and hence is critical. 

In brief, we have introduced a new analytically tractable model of
SOC. It is hoped that further studies of the model will contribute to 
a better understanding of self-organizing systems in general.

\centerline{\bf Figure Captions}

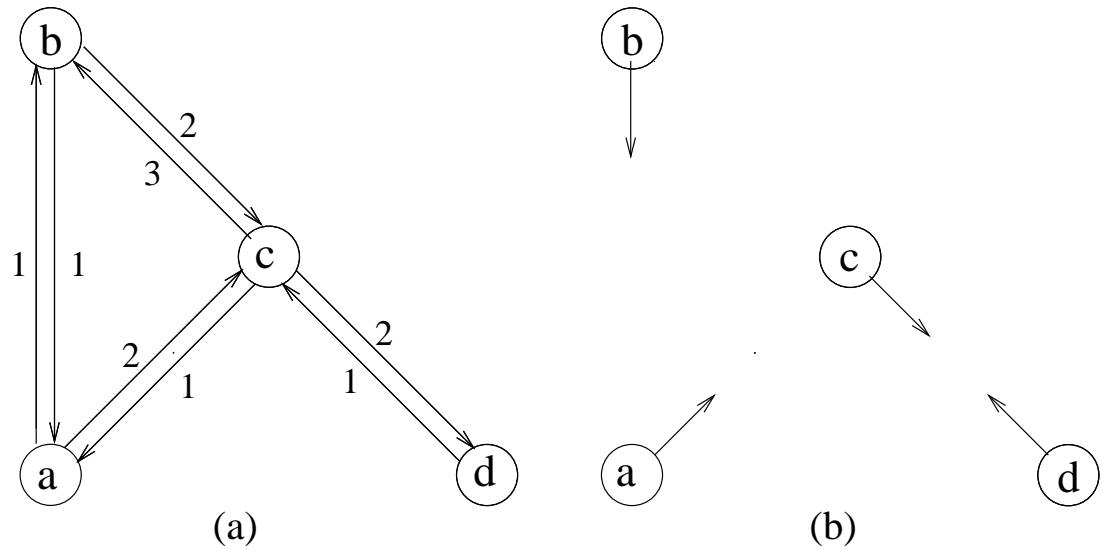
\begin{figure}
\caption{
(a) A directed graph. The outgoing bonds at each site are labelled 
by integers $1,2...$. An initial state with a
configuration of arrows as in (b) and a walker starting at the site
${\it{a}}$ moves along the path $abcb...$ which eventually settles to the
Euler circuit ${\it{abcdcbaca}}$.}
\end{figure}

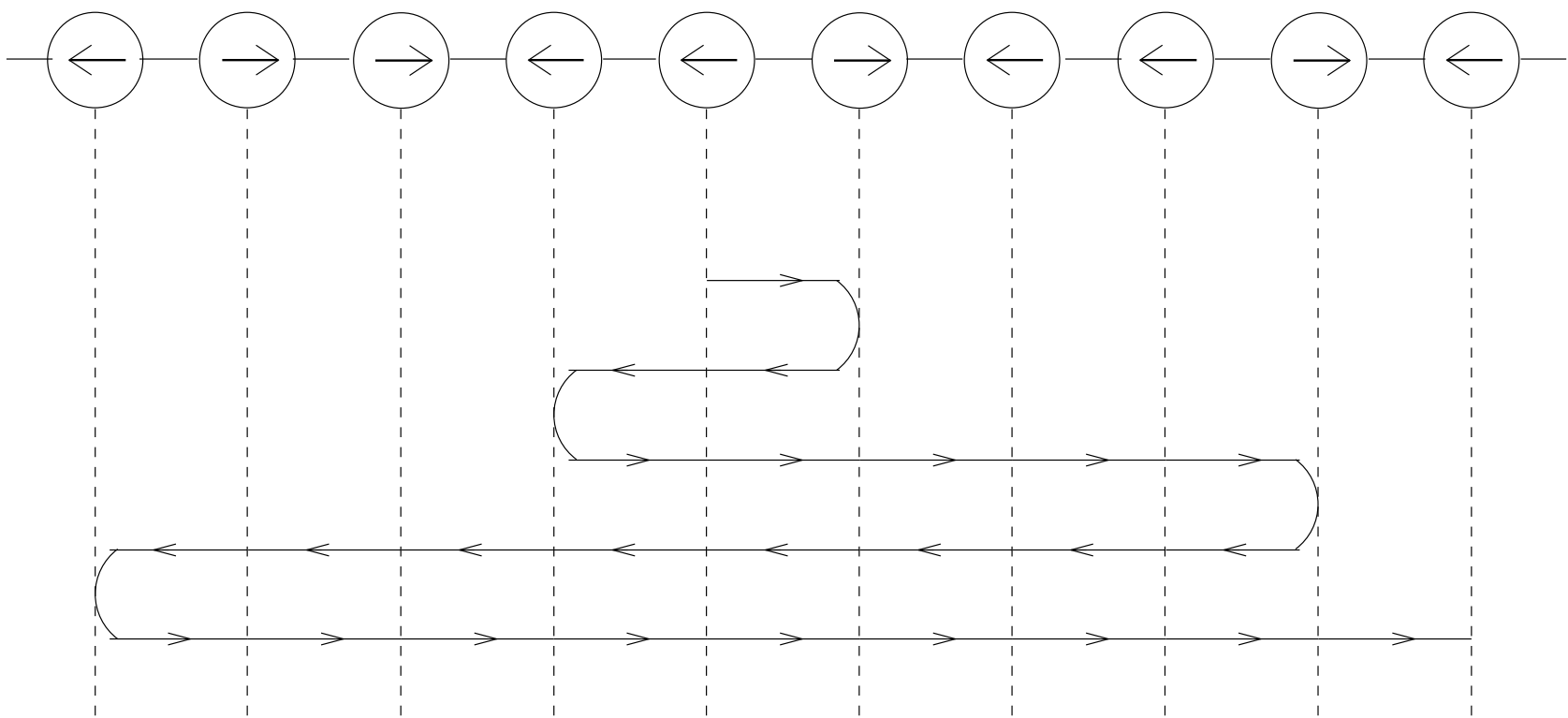
\begin{figure}
\caption{\label{fig2}
A random initial state of a lattice in one dimension and the motion of
a walker on this lattice. The medium is organized into a state in
which all arrows point in the same direction.}
\end{figure}

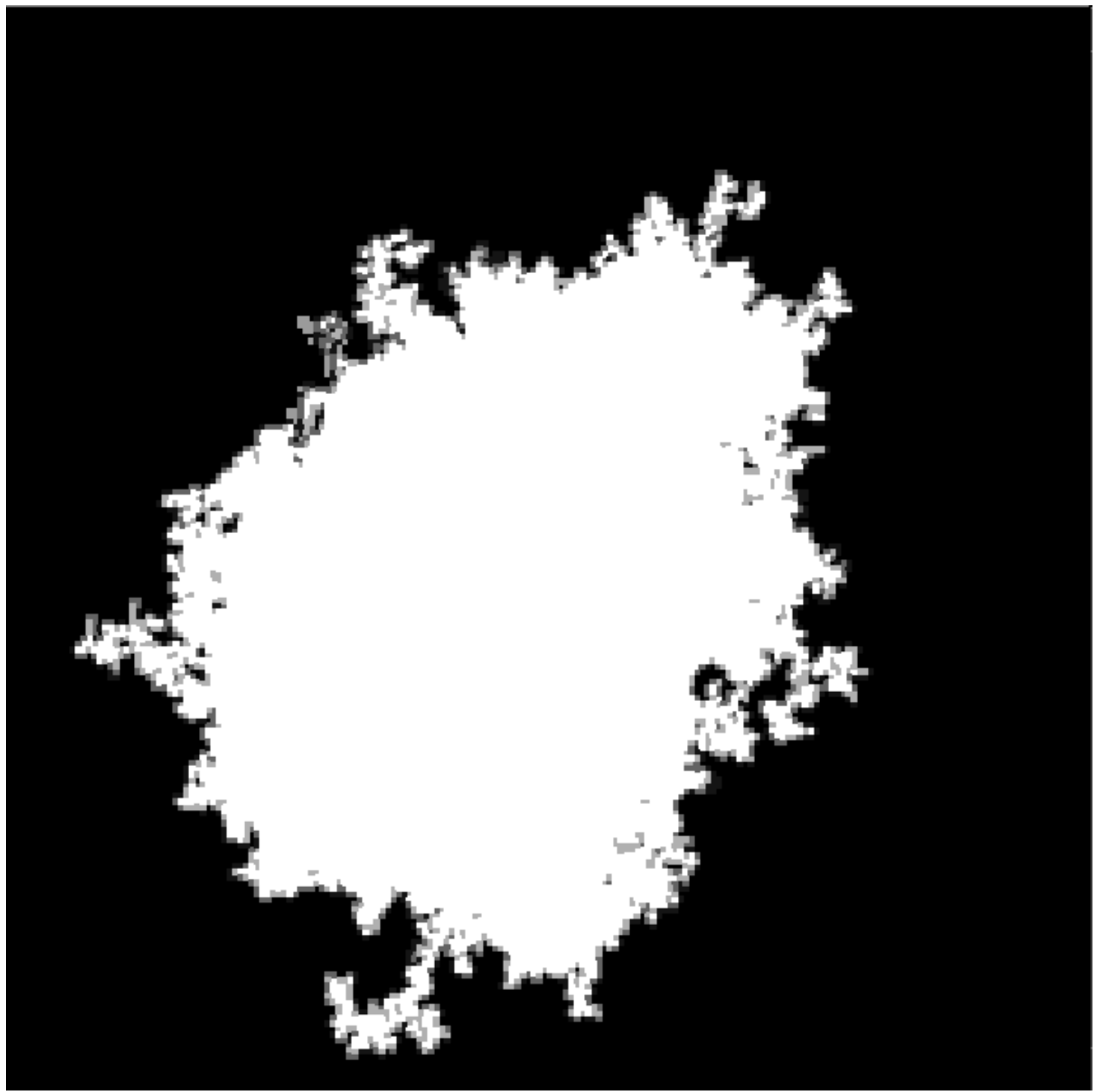
\begin{figure}
\caption{\label{fig3} Simulation of the Euler walk on a square lattice
with random initial conditions. The whole cluster consists of sites
covered by the walker after $10^5$ steps. The white
region shows the cluster of approximately $12500$ sites visited exactly
four times in the last $50,000$ steps.
The grey sites at the boundary of the cluster
are visited less than four times.}
\end{figure}

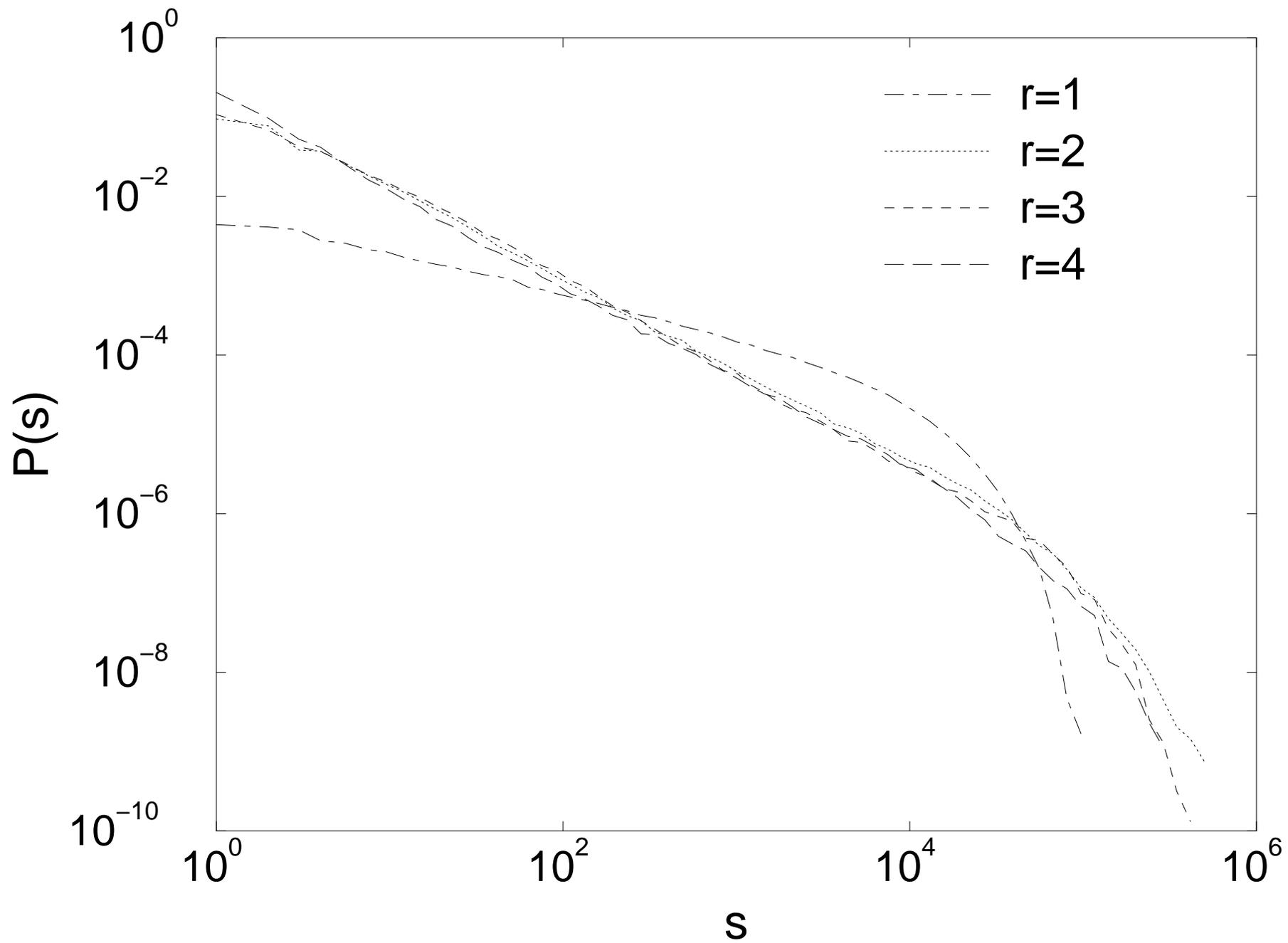
\begin{figure}
\caption{ \label{fig4}
 Plot of probability $P(s)$ of avalanche of size $s$ vs. $s$ for
different values of r.}
\end{figure}


\begin{thebibliography}{999}
\bibitem{note} The decay of correlations can be faster in special
directions if $\vec{e_1}$ or $\vec{e_2}$ are perpendicular to
$\vec{R_{12}}$ so that the coefficient of $R^{2-2d}$ vanishes.
\end{thebibliography}
\end{document}